\documentclass[lengthcheck,twocolumn,superscriptaddress,showpacs,10pt,prb,floatfix,amssymb]{revtex4}

\usepackage{graphicx}

\newcommand{\vsd}{V_\mathrm{SD}}

\newcommand{\iv}{$I$-$V$}

\newcommand{\Edi}{E_\mathrm{{D,i}}}
\newcommand{\Ed}{E_\mathrm{D}}
\newcommand{\Ef}{E_\mathrm{F}}
\newcommand{\Ec}{E_\mathrm{C}}
\newcommand{\Te}{\Theta_\mathrm{E}}
\newcommand{\Tc}{\Theta_\mathrm{C}}

\begin{document}

\newlength{\plotwidth}
\setlength{\plotwidth}{8.6cm}

\title{Shot noise in tunneling through a single quantum dot}

\author{A.~Nauen}
\affiliation{Institut f\"ur Festk\"orperphysik, Universit\"at Hannover,
Appelstr. 2, D-30167 Hannover, Germany}
\affiliation{Div. of Solid State Physics, Lund University, P.O. BOX 118, SE-221 00 Lund, Sweden}
\author{F.~Hohls}
\email{hohls@nano.uni-hannover.de}
\author{N.~Maire}
\affiliation{Institut f\"ur Festk\"orperphysik, Universit\"at Hannover,
Appelstr. 2, D-30167 Hannover, Germany}
\author{K.~Pierz}
\affiliation{Physikalisch-Technische Bundesanstalt, Bundesallee 100,
D-38116 Braunschweig, Germany}
\author{R.~J.~Haug}
\affiliation{Institut f\"ur Festk\"orperphysik, Universit\"at Hannover,
Appelstr. 2, D-30167 Hannover, Germany}

\date{\today}

\begin{abstract}
We investigate the noise properties of a zero-dimensional InAs quantum dot (QD) embedded
in a GaAs-AlAs-GaAs tunneling structure. We observe an approximately linear dependence of
the Fano factor and the current as function of bias voltage. Both effects can be linked
to the scanning of the 3-dimensional emitter density of states by the QD. At the current
step the shape of the Fano factor is mainly determined by the Fermi function of the
emitter electrons. The observed voltage and temperature dependence is compared to the
results of a master equation approach.
\end{abstract}
 \maketitle

The so called shot noise has been discussed initially for vacuum tubes,
where the current through the device fluctuates due to the stochastic
nature of the electron emission process.~\cite{schottky1918} A comparable
semiconductor device is a single tunneling barrier and the observed shot
noise follows the same expression as that in a vacuum tube: Its noise
power density $S=2eI$ is proportional to the average current $I$ with $e$
being the electron charge.~\cite{BlanterReview} However, it has been shown
that the amplitude of the shot noise for resonant tunneling through a
double-barrier structure is suppressed in relation to the so called {\it
Poissonian} value $2eI$. The occurrence of a suppression is independent
from the dimensionality of the resonant state: It has been observed for
the first time in quantum well structures where the tunneling takes place
through a 2-dimensional subband.~\cite{Li1990, Liu1995} Later experiments
in systems containing 0-dimensional states did also show a suppression of
the shot noise amplitude below the Poissonian
value.~\cite{birk1995,nauen2002} This suppression is caused by an
anti-correlation in the current due to the finite dwell time of the
resonant state in the tunneling structure.~\cite{Chen1991,Davies1992}.

In this paper we present noise measurements on self-assembled InAs quantum dot (QD)
systems. These samples provide ideal conditions for measuring the characteristics of
single 0-dimensional states since different individual QDs can be selected for transport
by applying different bias voltages between the source and drain contacts
\cite{narihiro1997, eaves1996, wurst2000}. In a previous paper~\cite{nauen2002} we
examined transport through an ensemble of quantum dots. Now we explore the regime of
transport through an individual quantum dot  in detail.

The active part of our samples consists of a GaAs-AlAs-GaAs resonant tunneling structure
with embedded InAs QDs of 10-15~nm diameter and 3~nm height.~\cite{wurst1999} These QDs
are situated between two AlAs barriers of nominally 4~nm (bottom) and 6~nm (top)
thickness. The thicker barrier is partially penetrated by the InAs QDs. This results into
an effective width of 3-4 nm which is slightly thinner than the bottom barrier. A 15~nm
undoped GaAs spacer layer and a GaAs buffer with graded doping on both sides of the
resonant tunneling structure provide three-dimensional collector and emitter electrodes.
Connection to the active layer is realized by annealed Au/Ge/Ni/Au contacts.

About one million QDs are placed randomly on the area of an etched diode structure of $40
\times 40\; \mu\mbox{m}^2$ area. However, it has been proven that only a small fraction
($\lesssim 1000$) of these QDs is actually able to participate in the electronic
transport.~\cite{wurst2003}

\begin{figure}[tb]
    \begin{center}
            \includegraphics[width=\plotwidth]{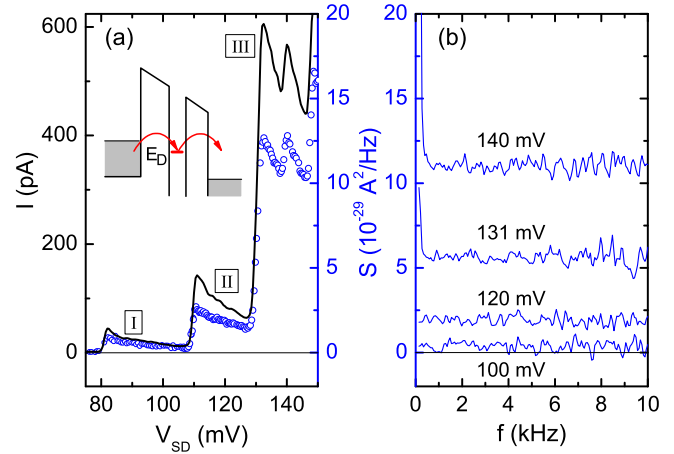}
    \end{center}
    \caption{%
        (a) Current-voltage characteristics of a GaAs-AlAs-GaAs double barrier tunneling
        structure with embedded InAs quantum dots at a temperature of $T=1.5$~K
        (black line, left axis) and shot noise amplitude $S$ as derived from
        averaging the curves in (b) for $f=1-10$~kHz (open symbols, right
        axis). The scale of the right axis was chosen such that the black line
        corresponds on this axis to the full Poissonian shot noise $S=2eI$
        expected for a single barrier structure.
        {\it Inset}: Schematic profile of the band structure at positive bias where
        resonant-tunneling through a QD is observed.
        \\
        (b) Typical noise spectra of the sample for different bias voltages.
        The data is smoothed with a 120~Hz boxcar average.
        The fluctuations of the signal
        increase with frequency due to the capacitive loading of the
        current amplifier.}
    \label{fig1}
\end{figure}

A diagram of the conduction-band profile with one InAs QD embedded in an AlAs barrier is
sketched in the inset of Fig.~\ref{fig1}a. Due to the small size of the InAs dots the
ground state energies $\Edi$ of all QDs are larger than the Fermi energy without applied
bias voltage. When applying a finite bias the zero-dimensional states of the QDs inside
of the AlAs barrier can be populated by electrons and a current through the structure
sets on. The largest quantum dots at the tail of the size distribution with lowest energy
are first getting into resonance. The small number of 'largest' dots adds the additional
selection for measuring transport through single InAs QDs.

A typical current-voltage ($I$-$V$) curve is shown in Fig.~\ref{fig1}a. We
observe a step-like increase of $I$ at bias voltages $\vsd
> 75$~mV. Each one of these current steps corresponds to the
emitter Fermi energy $\Ef$ getting into resonance with the ground states $\Edi$
of different individual QDs.

For positive bias voltages $\vsd > 0$ the electron tunnels first from the back contact
through the thicker bottom barrier onto the resonant state and then through the
effectively thinner barrier to the front contact. Thus the emitter tunneling rate $\Te$
is smaller than the collector tunneling rate $\Tc$ and the dot is mostly empty.
Therefore, the emitter tunneling rate dominates the current and allows us to study the
influence of the emitter on the noise properties.

For the noise measurements the sample is mounted into a specially crafted holder that
reduces the stray capacitance. This is necessary since the current noise is measured by a
low-noise current amplifier that tends to increase its internal noise in case of
capacitive loading. We used a current amplifier with bandwidth 10~kHz and inherent noise
level of nominally 10 $\mbox{fA}/\sqrt{\mbox{Hz}}$. The output signal is fed into a
Fast-Fourier-Transform analyzer for spectral decomposition. The sample holder itself is
installed in a $^4$He-cryostat with a variable temperature insert that can be flooded
with liquid helium.

In Fig.~\ref{fig1}b we show noise spectra for different applied bias voltages
after subtraction of the intrinsic amplifier noise and correction of the
amplifier gain. Frequency dependent $1/f$ noise appears only for high bias and
$f < 1$~kHz. For $f>1$~kHz we observe for the complete voltage range of
interest frequency-independent shot noise. We determine the shot noise
amplitude by averaging the spectrum from 1 to 10 kHz. The resulting voltage
dependence of the shot noise amplitude $S$ is shown by the open circles in
Fig.~\ref{fig1}a.

In order to characterize the amplitude of shot noise one usually compares
the measured values to the Poissonian value $2eI$ which is observed for
tunneling through a single barrier for $e\vsd \gg k_B T$. The scale of the
right axis in Fig.~\ref{fig1}a was chosen in such a way that the black
line corresponds to the full Poissonian shot noise. The comparison reveals
a suppression of the measured shot noise beneath $2eI$ which can be
understood as follows: As long as the ground state $\Ed$ of a QD is
occupied the tunneling of an additional electron from the emitter is
forbidden, resulting in an anti-correlation of successive tunneling events
on a timescale corresponding to the dwell time of the resonant state. This
makes the transport process less ``randomized" and consequently the shot
noise is reduced.\cite{nauen2002}

\begin{figure}[tb]
        \includegraphics[width=0.85\plotwidth]{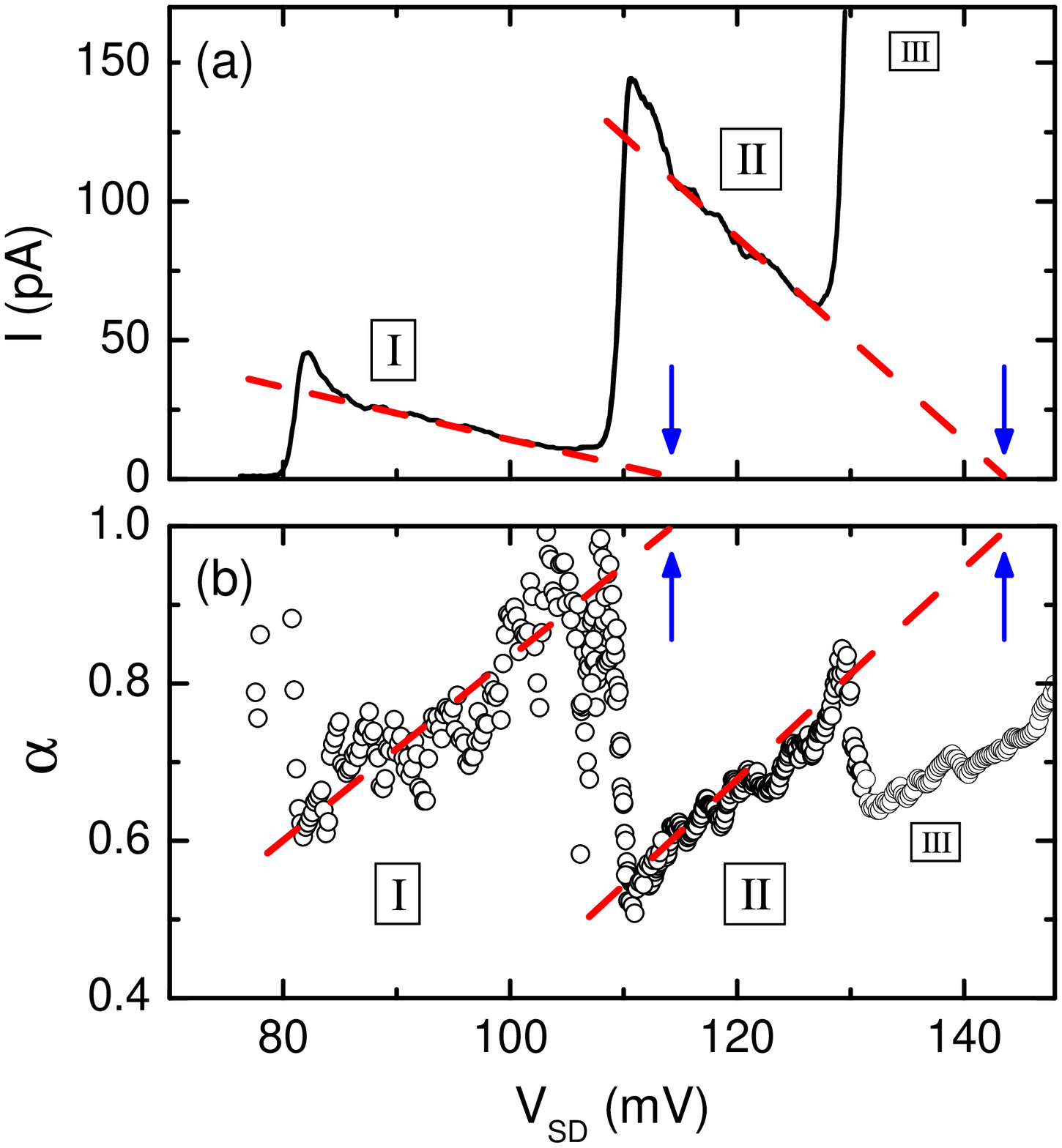}
    \caption{%
        (a) \iv characteristics for the transport through the first (I) and
        second(II)
        lowest lying resonance levels. Each corresponds to a different InAs dot
        in between the barriers. The dashed lines are guides to the eye to show
        the linear behavior of the current and its extrapolation to zero (see
        text).\\
        (b) Measured Fano factor $\alpha=S/2eI$
        of the InAs QDs.
        The data have been smoothed with
        a 5-point boxcar average.
        Again the dashed lines are guides to the eye for the linear behavior.
        }
    \label{fano}
\end{figure}

We will now concentrate onto the two well resolved steps at $\vsd=80$~mV and 110~mV,
denoted with (I) and (II). Fig.~\ref{fano}a focuses onto this part of the \iv-curve. With
increasing voltage $\vsd$ the dot energies $\Edi = \Edi^0 - \beta e \vsd$ are lowered
with respect to the emitter (lever arm $\beta \approx 0.4$). For each resonance level
crossing the Fermi energy from empty to occupied emitter states we observe first a step
like increase of the current. With further decreasing energy $\Edi$ the current drops
linearly as indicated by the dashed lines. This nicely matches the prediction of Liu and
Aers~ \cite{liu1988,liu1989} for resonant 3d-0d-3d tunneling.

The observed linear decrease of the current is related to the scanning of the
density of states (DOS) of the 3-dimensional emitter by the QD ground state. We
can describe this in terms of energy resp.\ voltage dependent tunneling rates
$\Theta_\mathrm{E,C}(\vsd)$ of emitter and collector. Neglecting the energy
dependence of the wave function overlap \cite{liu1989} the tunneling rates are
proportional to the area $A(\Ed)\propto \Ed-\Ec$ in momentum space satisfying
energy conservation.~\cite{liu1988} Thus $\Theta(\Ed)\propto \Ed-\Ec$ depends
linearly on distance of the dot energy $\Ed$ to the conduction band edge $\Ec$.
Assuming $\Te \ll \Tc$ due to the asymmetric barriers the current $I \approx 2
e \Te(\vsd)$ acquires the observed linear dependence.

For our sample with a Fermi energy $\Ef \approx 14$~meV and an energy-to-voltage
conversion factor of $\beta \approx 0.4$ the current falls back to zero when the distance
to the onset voltage exceeds $\Delta V \approx 35$~mV, since then the QD ground state
with energy $\Edi$ moves below the conduction band edge $\Ec$ of the emitter. This agrees
with the extrapolation of the current plateau by the dashed lines towards $I=0$ in
Fig.~\ref{fano}a.

The afore mentioned approximate linear dependence of the current is mirrored in the
behavior of noise properties. In Fig.~\ref{fano}b we plot the Fano factor $\alpha$
defined as the ratio $\alpha=S/2eI$ of the measured noise $S$ to the full Poissonian shot
noise $2eI$. At the step edges of the current we observe maximal noise suppression resp.\
minimal $\alpha$. With further increase of $\vsd$ the Fano factor rises approximately
linearly until the next quantum dot comes into resonance. In a previous
experiment~\cite{nauen2002} the quick succession of new QDs lead to the observation of a
series of peaks in $\alpha$. In the present sample the large spacing of the quantum dot
energies allows to observe and extrapolate the linear dependence of the Fano factor for a
single resonance. We find a value $\alpha\approx 1$ for the same $\vsd$ value at which
the current vanishes.

We can calculate the expected Fano factor using a master equation approach
following Kiesslich {\sl et al.}~\cite{kiesslich2002,kiesslich2003} For a
spin degenerate ground state and forbidden double occupancy due to Coulomb
blockade we find
\begin{equation}
    \alpha=1-\;\frac{4  \Te \,\Tc}
        {\left(2\Te+\Tc\right)^2}
        \approx 1-4\frac{\Te(\vsd)}{\Tc} +
        \mathrm{O}\!\left({\scriptstyle\frac{\Te^2}{\Tc^2}}\right).
    \label{f1}
\end{equation}
Here we have set $f_E(E)=1$ and $f_C(E)=0$ for the emitter and collector
Fermi functions. In the second step we kept only terms of order $\Te/\Tc$.
We also omit the voltage dependence of $\Tc$ as it changes only weakly in
the relevant $\vsd$-window: Due to the large bias voltage the electrons
tunnel into collector states at energies high above the Fermi energy and
the conductance band edge. The change in the collector tunneling rate is
only of order $\Ef/e\vsd\approx 0.1$ for a change of $\vsd$ from the step
edge to vanishing current for a single resonance.

With Eq.~\ref{f1} we easily understand that the linear behavior of the
Fano factor has the same origin as the linear behavior of the current,
namely the linearly vanishing tunneling rate $\Te(\vsd)\propto V_0-\vsd$
with $V_0$ the voltage at which $\Ed$ crosses $\Ec$. Near this point
$\Te\ll\Tc$ and we observe essentially single barrier tunneling with full
Poissonian shot noise $S=2eI$ and thus $\alpha = 1$.

The smallest value of the Fano factor of $\alpha \approx 0.55$ shows up at the current
step edge of QD (II) in Fig.~\ref{fano}. Following Eq.~\ref{f1} this corresponds to an
asymmetry of the tunneling rates $\Tc/\Te\approx 4$. In case of QD (I) the asymmetry is
increased since the maximal suppression is $\alpha\approx 0.62$, corresponding to
$\Tc/\Te\approx 6$. This difference most likely stems from the height distribution of the
InAs QDs resulting in differing effective thicknesses of the collector barrier.

\begin{figure}[tb]
        \includegraphics[width=\plotwidth]{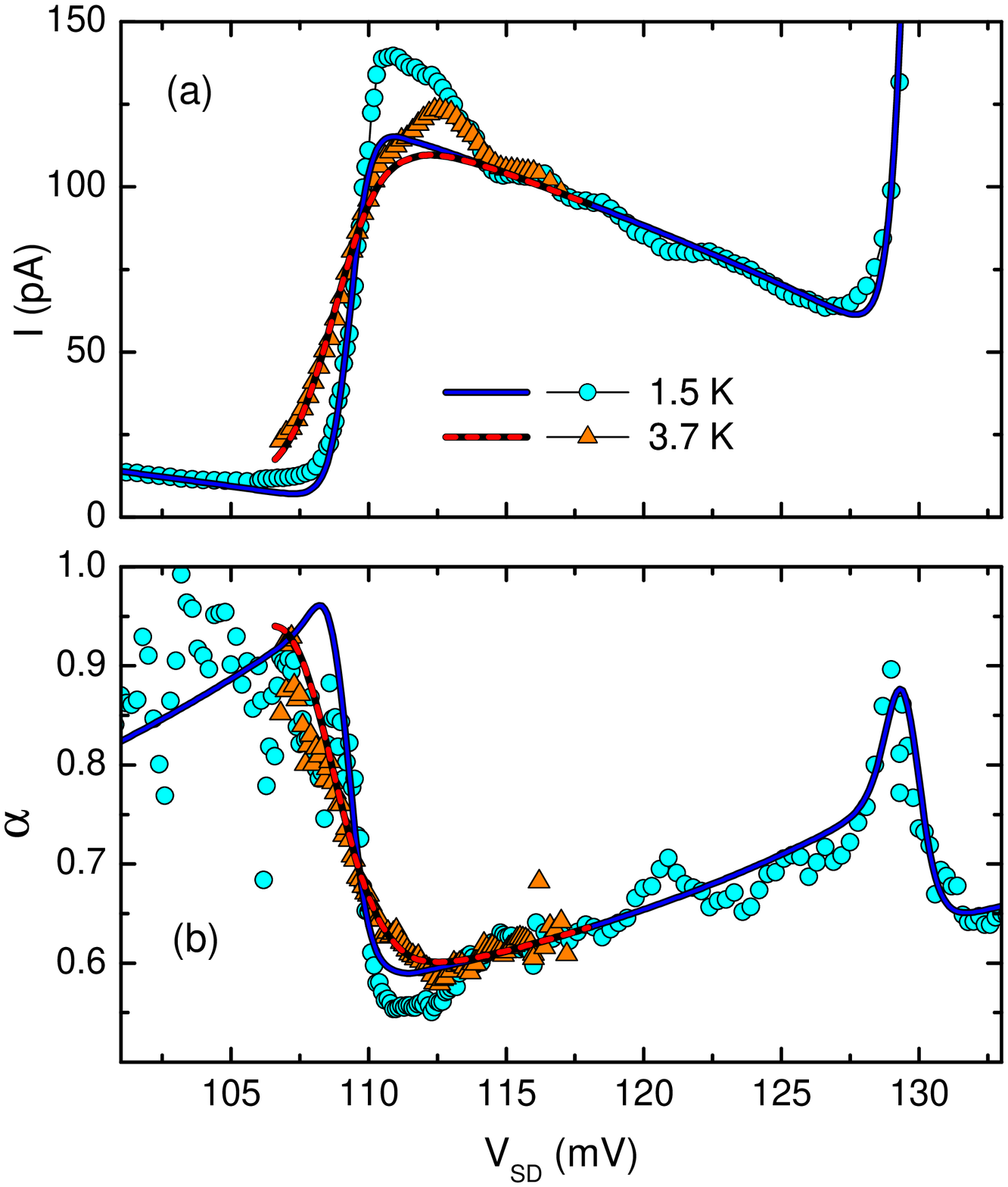}
    \caption{
        (a) Modelling of the current $I$ for transport through quantum dot (II).
        Symbols denote measured data, lines are the result of the model
        (Eq.~\ref{f-strom}), extended with contributions of QD (I) and (III). The
        comparison of two different temperatures demonstrates the softening of the step
        edge due to the Fermi distribution.
        (b) Same for the Fano factor $\alpha$, which is modelled by Eq.~\ref{f-fano}.
        The data for the Fano factor have been smoothed with a 5-point boxcar average.
        }
    \label{simu}
\end{figure}

We will now concentrate our analysis onto the temperature dependence of
transport through QD (II) which yields a larger current and thus a
stronger noise signal. Fig.~\ref{simu} displays the measured current and
Fano factor for two different temperatures. The temperature affects mainly
the step edge: When shifting the resonant level $\Ed$ through the Fermi
energy $\Ef$ the current changes smoothly due to the finite width of the
Fermi function. In a first approximation this could be modelled by a
voltage and temperature dependent tunneling rate $\Te^T(T,\vsd)\propto
f_E\left(T,\Ed(\vsd)\right)A\left(\Ed(\vsd)\right)$ with $A(\Ed)$ the area
in momentum space as described above and
$f_E^{-1}=1+\exp\left(\left(\Ed(\vsd)-E_F\right)/k_B T\right)$ the Fermi
function. In this descriptive approach the tunneling rate is proportional
to the occupied density of states fulfilling energy conservation. For a
more rigid evaluation we use a master equation
approach~\cite{kiesslich2002,kiesslich2003} which yields the following
formulas for the current $I$ and the Fano factor $\alpha$:
\begin{equation}
     I=\frac{2 e f_E\;\Te(\vsd)\Tc}
        {\left(1+f_E\right)\Te(\vsd)+\Tc} \;,
     \label{f-strom}
\end{equation}
\begin{equation}
    \alpha=1-\;\frac{4 f_E\; \Te(\vsd) \Tc}
        {\left(\left(1+f_E\right)\Te(\vsd)+\Tc\right)^2}\;.
    \label{f-fano}
\end{equation}
The equations were derived for a spin degenerate quantum dot with forbidden double
occupancy due to Coulomb energy.

In order to fit the theoretical expression for the current $I(\vsd)$ (Eq.~\ref{f-strom})
and the Fano factor $\alpha(\vsd)$ (Eq.~\ref{f-fano}) to the experimental data we use the
following procedure: The ratio of the tunneling rates $\Tc/\Te=4.4$ at the step edge is
deduced from the Fano factor while the absolute value $\Tc=2.4 \cdot 10^9~1/\mbox{s}$ is
gained from a fit of the current. The lever arm $\beta=0.35$ and the dot energy were
chosen for a best match of the step edge, and the linear extrapolation of $\Te$ regarding
to the scanning of the emitter DOS was fitted to the further evolution of the Fano factor
on the current step. For best agreement we include the contributions of QD (I) and QD
(III) which are relevant at the onset of current through QD (II) and for $\vsd > 127$~mV
where transport through QD (III) sets in. As described in Ref.\cite{nauen2002} for
transport through multiple dots we use $I=\sum I_i$ and $\alpha = \sum (I_i/I)\alpha_i$
with $I_i$ and $\alpha_i$ for each dot given by equations \ref{f-strom} and \ref{f-fano}.

In Fig.~\ref{simu} we show the results of the afore discussed procedure for current $I$
and Fano factor $\alpha$ in comparison to the experimental data for two different
temperatures. It is evident that both the current through QD (II) and the corresponding
noise suppression can be described satisfactorily by the above sketched model. Also the
peak in the Fano factor at $\vsd \approx 128$~mV is well described by the sum of
contributions from QD (II) and QD (III) which confirms Ref.~\cite{nauen2002} where
several peaks in $\alpha$ were observed at each onset of current through an additional
quantum dot.

This dependence of the Fano factor $\alpha$ on $\vsd$ underlines
unambiguously that the suppression of the shot noise is indeed
linked to the ratio of the tunneling rates since due to the
3d-0d-3d tunneling in our experiment we are able to tune $\Te$ in
relation to $\Tc$. Furthermore we can conclude that the crossover
from full shot noise $\alpha=1$ into the suppression is, just as
the step edge of the current, governed by the Fermi function $f_E$
of the emitter.

Nevertheless, we observe some deviations from our simple model and we will discuss these
now: We observe some fine structure in the current and the shot noise in the 'plateau'
region which is caused by the fluctuations of the local density of states of the
emitter.~\cite{schmidt2001}

However, the main difference between the experiment and the theoretical model
is the overshoot of the current $I$ directly after the step edge for
$110~\mbox{mV}\lesssim \vsd \lesssim 113~\mbox{mV}$. Most probably this is
related to a Fermi-edge singularity (FES) that was shown to enhance the
tunneling near the threshold when the QD state is resonant with the emitter
Fermi energy. It is caused by a Coulomb interaction of the fluctuating charge
on the QD and the emitter electron reservoir.~\cite{wurst2000,geim:prl:1994}
Interestingly the Fano factor does also reveal a stronger shot noise
suppression below the value given by Eq.~\ref{f-fano}. For increased
temperature the overshoot of the current $I$ has virtually vanished as expected
for a FES effect.~\cite{wurst2000,geim:prl:1994} Also the additional
suppression of the Fano factor below the single particle expectation
(Eq.~\ref{f-fano}) has vanished. Therefore we assume that both features are
caused by the same physical process, that is electron-electron interaction.


To conclude, we have measured the shot noise suppression for resonant 3d-0d-3d tunneling
through a single InAs QD. We could show that the Fano factor $\alpha$ is linked to the
ratio of the tunneling rates through emitter and collector barrier, $\Te$ and $\Tc$
respectively. We model the observed voltage and temperature dependence of current and
shot noise following a master equation approach and find in general a good agreement.

The authors would like to thank Gerold Kiesslich for enlightening
discussions.

We acknowledge financial support from DFG and BMBF.


\bibliographystyle{apsrev}

\end{document}